\documentclass[fleqn,12pt]{article}

\usepackage{espcrc1}
\usepackage{graphicx}

\usepackage[figuresright]{rotating}
\title{Ground-state correlations and final state interactions in exclusive lepton
scattering off few-nucleon systems\footnote{Presented at the {\it Third
International Conference on Perspective in Hadronic Physics}, Trieste,
 7-11 May, 2001. To appear in  {\it  Nuclear Physics}.}}

\author{C. Ciofi degli Atti
\address{Department of
 Physics, University of Perugia and
 INFN, Sezione di Perugia,
 via A. Pascoli, Perugia, I-06100, Italy}, L. P. Kaptari$\, ^{\rm a}$\thanks
{On leave from  Bogoliubov Lab. Theor. Phys.,
        JINR, Dubna, Russia}}

\begin{document}
\maketitle

\begin{abstract}
  The two nucleon emission process off $^3He$ induced by medium energy electrons
  has been theoretically analyzed using realistic
  three-nucleon wave functions and taking the final state interaction into account.
  Various kinematical conditions have been considered in order to
  clarify the question  whether the effects
  of  the final state interaction could be  minimized by a proper choice of the
  kinematics.
\end{abstract}

\def \bfgr #1{ \mbox {{\boldmath $#1$}}}
\newcommand{\be}{\begin{eqnarray}}
\newcommand{\ee}{\end{eqnarray}}
\newcommand{\nn}{^3He(e,e'N_2N_3)N_1}
\newcommand{\pp}{^3He(e,e'p_1p_2)n}
\newcommand{\pn}{^3He(e,e'p_1n)p_2}
\newcommand{\ga}{\gamma^*}
\newcommand{\emme}{M({\bf p}_m,{\bf t})}
\newcommand{\enne}{M({\bf k}_1,{\bf t})}
\newcommand{\spectral}{P_1(k_1,E^*)}

\section{Introduction: correlations and final state interactions in lepton scattering
off nuclei}
The  investigation of Ground State  Correlations (GSC) in nuclei, in particular those which
originate from the most  peculiar features of the Nucleon-Nucleon (NN)  interaction,
i.e. its strong short range repulsion  and  complex state  dependence (spin,
isospin, tensor, etc),
is one of  the most challenging aspects of experimental and theoretical
nuclear physics and, more generally, of hadronic physics. The results of sophisticated
many-body calculations in terms of realistic models of the NN interactions, show
 that the complex structure of the latter generates a rich correlation structure of the
 nuclear ground state wave function,  but the investigation of such a structure is problematic
 due to the effects of the final state interaction (FSI); these, in fact,
  very often compete  with  the
effects generated by GSC, so that the  longstanding question
{\it Does FSI hinder the investigation of GSC?}
  has not yet been
 clearly answered.  To-day the answer could  probably  be
  provided
 in a more reliable way, particularly in the case of few-body systems, for which accurate
 ground state wave functions are available and  FSI effects can also be calculated in a
 satisfactory way (see e.g. \cite{gloeckle}, \cite{rosati}). In this paper
 the process of
  two-nucleon emission off the three-nucleon systems, which is under
  intense experimental investigation (see e.g. \cite{eddy},\cite{larry})  will be discussed, with the aim of providing
 another attempt  at answering
  the above longstanding question.
 Realistic three-nucleon wave functions \cite{rosati} corresponding to the AV18
 interaction \cite{AV18}, will be used, and the effects from the FSI will
 be investigated at various levels of complexity.

 \section{Two-nucleon emission off the three-nucleon systems}
 We  consider the absorption of a virtual photon ${\gamma^*}$ by  a nucleon
  bound in $^3He$ followed
 by two-nucleon emission, in particular the process $\nn$, where
  $N_2$ and $N_3$ denote
 the two nucleons which
  are detected and  $N_1$ the third one. In what
  follows $Q^2={\bf q}^2-{\nu}^2$ denotes  the photon four-momentum
   transfer, $\bf k_i$
 the  momenta of the bound nucleons before ${\gamma^*}$ absorption,
 and $\bf p_i$  the momenta in the continuum
 final state.

 Momentum and energy conservation  require that

 \be
 {\sum_{i=1}^3{\bf k}_i} =0, \qquad\qquad  {\sum_{i=1}^3{\bf p}_i = {\bf q}},
 \qquad\qquad  \nu +M_3 =\sum_{i=1}^3 (M^2+{\bf p}_i^2)^{1/2}
 \label{one}
 \ee
 where $M$ and $M_3$ are the nucleon and the three-nucleon system masses, respectively.

 In one-photon exchange approximation the cross section of the
process reads as follows
\be
\frac{d^{12}\sigma}{d\epsilon^{'} d\Omega{'} d{\bf p_1} d{\bf p_2} d{\bf p_3}}={\sigma}_{Mott}\cdot
\sum_{j=1}^6 v_j \cdot W_j \cdot \delta( {\bf q} - \sum_{i=1}^3{\bf p}_i ) \delta( \nu +M_3 -\sum_{i=1}^3 (M^2+{\bf p}_i^2)^{1/2})
\label{three}
\ee
where $v_j$ are well known kinematical factors,  and $W_j$  the {\it response functions},
which  have the following general form
\be
W_j \propto\left | \langle \Psi_f^{(-)}({\bf p}_1, {\bf p}_2, {\bf p}_3)
|\hat{ \mathcal O}_j({\bf q})|
\Psi_i({\bf k}_1, {\bf k}_2, {\bf k}_3\rangle \right |^2
\label{four}
\ee
In Eq. (\ref{four}) $|\Psi_f^{(-)}({\bf p}_1, {\bf p}_2, {\bf p}_3)\rangle$ and
$|\Psi_i({\bf k}_1, {\bf k}_2, {\bf k}_3)\rangle$
are the continuum and ground state wave functions of the three body system, respectively,
and   $\hat{ \mathcal O}_j({\bf q})$ is a  quantity depending on proper combinations of the components
of the  nucleon
current  operator  $\hat{ j^{\mu}}$ (see e.g. \cite{report}).


\begin{figure}[h] 
\begin{center}
    \includegraphics[height=0.50\textheight]{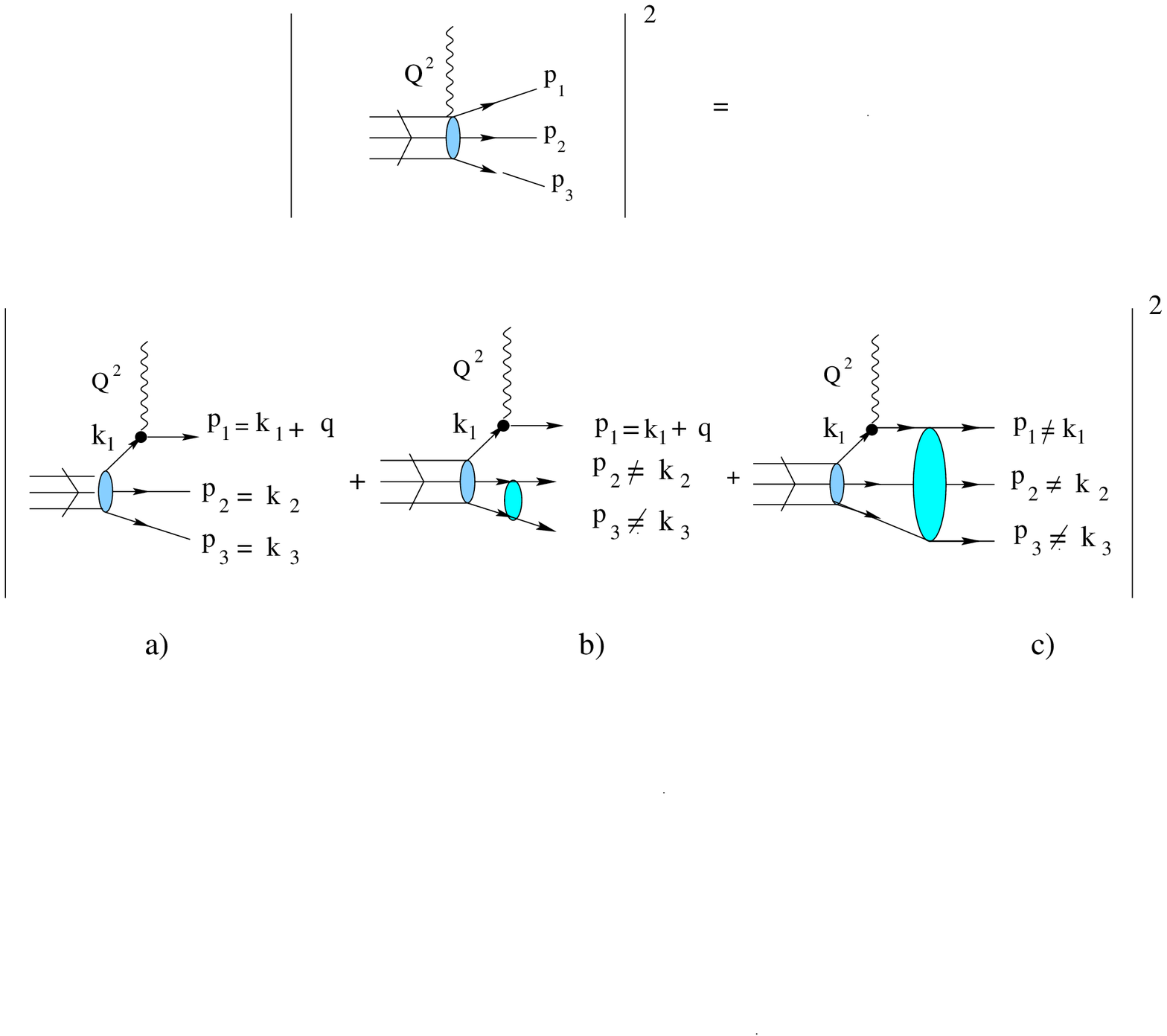}
\vskip -4cm
\caption{ The various processes contributing to  the
 reaction  $\nn$: (a) {\it No FSI}, (b)  the {\it NN rescattering},
  (c) the {\it  three-body rescattering}.
 Note that in this paper, following Ref. \cite{CPS},  the sum of $a)$ and $b)$ is called
 {\it The Plane Wave
 Impulse Approximation (PWIA)}, whereas in Ref. \cite{gloeckle}
  PWIA is the same as our {\it No FSI}
 (process $a)$).}
\label{fig2}
\end{center}
\end{figure}
\noindent The various processes, in order of increasing complexity,
 which contribute to the reaction $^3He(e,e'N_2N_3)N_1$ are depicted in
 Fig. \ref{fig2}.

  Let us introduce
 the {\em relative},  ${\bf t}=\displaystyle\frac{{\bf p}_2-{\bf p}_3}{2}$,
 and  {\em Center-of-Mass},
 ${\bf P}={\bf p}_2+{\bf p}_3$, momenta of  the detected
 pair,  and
  the {\it missing momentum}
${\bf p}_m={\bf p}_1-{\bf q}= -({\bf p}_2+{\bf p}_3)$.
As already stated, we consider the process  $\pp$ ( $\pn$) , in which $\ga$
 interacts with the neutron (proton) and the two protons (proton-neutron) {\it correlated in the
 initial state} are emitted and detected. Within the PWIA, i.e.  when  the final state rescattering
 between the two detected nucleons is taken into account  (processes  $a)$ and $b)$), but
 the interaction of the hit neutron(proton) with the emitted proton-proton
 (proton-neutron) pair is disregarded,
 the  cross section (Eq. (\ref{three})) integrated
 over ${\bf P}$ and the kinetic energy of $N_1$, has the following form
 (we take  ${\bf q }\parallel  z$)
\be
\frac{d^{8}\sigma}{d\epsilon^{'} d\Omega{'} d\Omega_{N_1} d{t}  d\Omega_t}
&&\!\!\!\!\!={\mathcal K}\left( Q^2,\nu,{\bf p}_m,{\bf t}\right)
\,\cdot\,\frac12
\sum\limits_{M_3,\sigma,s_f,\mu_f}
\left|\int \exp( {i {\bf p}_m \bfgr\rho})
\, \chi_{\frac12 \,\sigma}
\Phi_{s_f\mu_f}^{{\bf t}(-)}({\bf r})
 \Psi_{3M_3}^*({\bf r},{\bfgr \rho})\right |^2
\nonumber\\
&&\!\!\!\!\!\equiv {\mathcal K}\left(Q^2,\nu,{\bf p}_m,{\bf t}\right)\cdot M({\bf p_m}, {\bf t})
\label{nine}
\ee
where  ${\bf p}_m= - {\bf k}_1$, ${\mathcal K}$
 incorporates all kinematical factors,
 $\chi_{\frac12 \,\sigma}$ represents the Pauli spinor for the hit particle,
$\Phi_{s_f\mu_f}^{{\bf t}(-)}(\bf r)$ is the two-nucleon wave function in the
continuum, and ${\bf r}$ and ${\bfgr \rho}$ are usual Jacobi coordinates.

 In the rest of the paper we will omit, for ease of presentation,
all explicit summations over the quantum numbers and will denote the
continuum two-nucleon wave function simply by  $\Phi_{N_2N_3}^{{\bf t}^{(-)}}$.
The quantity $M({\bf p}_m, {\bf t})$ can then be cast in the following simple
form
\be
&&
M({\bf p}_m, {\bf t})= M({\bf k}_1, {\bf t})=  \left|\int \exp( {i {\bf p}_m \bfgr\rho})
\, I_{N_2N_3}^{\bf t}({\bfgr \rho})d^3\rho \right|^2
\label{ten}
\ee
where   $I_{N_2N_3}^{\bf t}({\bfgr \rho})$ is the {\it overlap  integral}
between the three-nucleon ground state wave function and the two-nucleon continuum state,
 i.e.
\be
&&
 I_{N_2N_3}^{{\bf t}}({\bfgr \rho})=\int\Phi_{N_2N_3}^{{\bf t}^{(-)}} ({\bf r})\Psi_{3M_3}({\bf r},
  {\bfgr \rho})d^3 r
\label{eleven}
\ee

Within the {\it No FSI} approximation,i.e. when only process $a)$ contributes to the
reaction,   one has
$\Phi_{N_2N_3}^{{\bf t}^{(-)}}\propto \exp({i{\bf t} \cdot {\bf r}})$ and
$
 I_{N_2N_3}^{{\bf t}}({\bfgr \rho})=\int e^{i {\bf t}{ \bf r}}
 \Psi_{3M_3}({\bf r}, {\bfgr \rho})d^3 r
$,
so that
\be
&&
M({\bf k}_1, {\bf t})=\left| \int \exp( {i {\bf k}_1 {\bfgr {\rho}}}) \exp({i {\bf t}{ \bf r}})
\Psi_{3M_3}({\bf r},
 {\bfgr \rho})d^3 r
d^3\rho \right|^2
\label{4teen}
\ee
 represents nothing but the square of the three-nucleon wave function in momentum space.
\begin{figure}[h] 
\begin{center}
    \includegraphics[height=0.5\textheight]{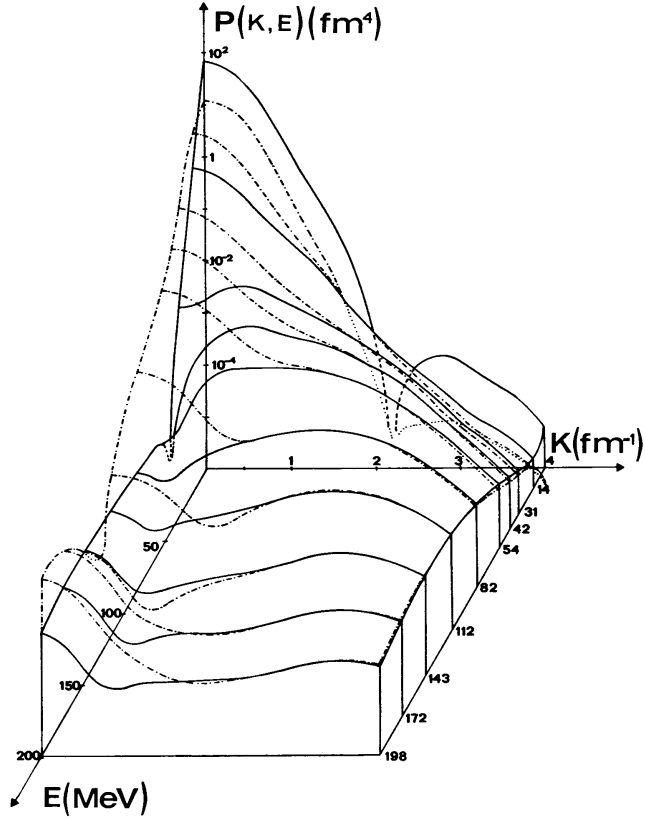}
\caption{The three-body channel neutron (proton) Spectral Function
in $^3He$ ($^3H$) ($k\equiv k_1 $). The dot-dashed line
corresponds to the {\it No FSI} case, whereas the full line includes the
 neutron-neutron (proton-proton)  rescattering. Three-nucleon
 wave function from \cite{CPS}; Reid Soft Core interaction
\cite{Reid}
(After Ref. \cite{CPS}).}
\end{center}
\label{fig3}
\end{figure}
When the PWIA is considered (process $a)$ plus process
$b)$), the direct correspondence between the three-nucleon wave function and
the cross section is lost and the process is governed by the quantity
$M({\bf k}_1,
 {\bf t})$ provided by Eq. \ref{ten}. The integral of the latter  over the direction of ${\bf t}$,
 is related to the {\it Spectral Function} of  nucleon
$N_1$, namely
\be
&&
\frac{M|{\bf t}|}{2}\, \int M({\bf k}_1, {\bf t}) d\Omega_{\bf t}=P_1(k_1,E^*)
\label{5teen}
\ee
where
$ E^*={\bf t}^2/M$
 is the "excitation energy" of the spectator pair $N_2N_3$, which is related to the {\it removal energy} $E$
 of nucleon $N_1$ by
 $
 E= E_3 + E^* $,
 where $E_3$ is the (positive) binding energy of the three nucleon system.
If the Coulomb
 interaction is disregarded, the neutron Spectral Function in$^3He$
 is the same as the proton Spectral Function in $^3H$.
 In Fig. \ref{fig3} the nucleon
 Spectral Function calculated with and without the {\it NN rescattering} is shown
 \cite{CPS}.
  It can be seen that
 there is a  region where the  FSI ({\it NN rescattering})
  does not play any role. This is the so called
 {\it  two-nucleon correlation} region, where the relation
$ E^* ={\bf t}^2/M \simeq {{\bf k}_1}^2/4M$
 holds (see e.g. Ref. \cite{2NC}). The existence  of such a region is a general feature of any
 Spectral Function, independently of the two-nucleon interaction and of  the method to generate the
 wave function. This is illustrated in Fig. \ref{fig4},  where the Spectral
 Function obtained with the variational wave function  of Ref. \cite{rosati},
 corresponding to  the $AV18$ \cite{AV18} interaction, is shown for
 several  values of $k \equiv k_1$.

  From the figures we have exhibited  one expects that if the kinematics is properly chosen
 (i.e. $E^* = {\bf t}^2/M \simeq {{\bf k}_1}^2/4M$) the {\it  NN rescattering}
  can be strongly
 reduced;  on the contrary, if  it is chosen improperly (in particular corresponding to
 an initial state characterized by
  $k_1 \simeq 0$),
  the  {\it NN rescattering} fully distorts the {\it No FSI}  predictions. This is
  demonstrated
 in Fig. \ref{fig5} where
  Eq. (\ref{ten}) is shown and
  compared with the {\it No FSI} approximation (Eq. (\ref{4teen})).
 In this calculation,
  we have fixed the two-nucleon relative energy
  $E^*={\bf t^2}/M=50 MeV$ and have plotted, for a  given values of  $k_1$,
   the dependence of $\enne$ upon the angle $\theta_{\bf t}$ between $\bf t$ and $\bf q$,
   the latter being  chosen along
   ${\bf k}_1$. The four values of $k_1$  correspond to  three
    relevant regions of the
   Spectral Function, {\it viz}  :
   1. $E^* >  \frac{{\bf k}_1^2}{4M}$ (${k}_1=0.5 fm^{-1}$ and  $1 fm^{-1}$);
   2.  $E^* \simeq \frac{{\bf k}_1^2}{4M}$ (${k}_1=2.2 fm^{-1}$), the {\it
   correlation region};
   3.  $E^* <  \frac{{\bf k}_1^2}{4M}$ (${k}_1=3 fm^{-1}$).

\vskip 4mm
 \begin{figure}[h] 
\begin{center}
    \includegraphics[height=0.40\textheight]{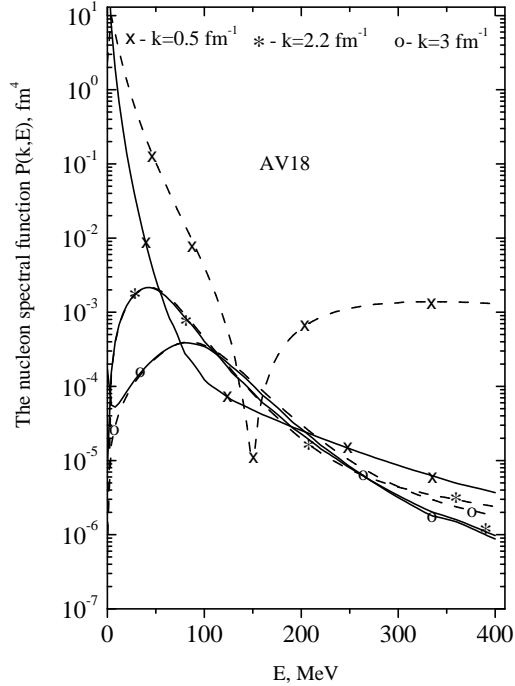}
\caption{The Nucleon Spectral Function as in Fig.\ref{fig3} corresponding to the wave function
of Ref. \cite{rosati} and the $AV18$ interaction \cite{AV18} ($k\equiv k_1$).
 Dashed line: {\it No FSI}; full line: {\it NN rescattering}.}
\label{fig4}
\end{center}
\end{figure}
It can be seen that in the first region the two nucleon rescattering is very large (cf. Figs.
\ref{fig3} and \ref{fig4}), whereas in the two other regions, it is very small.

We have also investigated the effect of $NN$ rescattering on a particular kinematics, namely that
which corresponds to the initial state in which $N_2$ and $N_3$ are correlated with momenta
${\bf k}_2$ = -${\bf k}_3$  and  ${\bf k}_1$ = $0$, so that,
 after $\ga$ absorption,
 $N_1$ is emitted  with
momentum $\bf q$,  and  $N_2$ and $N_3$ are emitted  back-to-back  with momenta
${\bf p}_2$ = -${\bf p}_3$ (${\bf p}_m = 0$).
\begin{figure}[h] 
  \begin{center}
    \includegraphics[height=0.4\textheight]{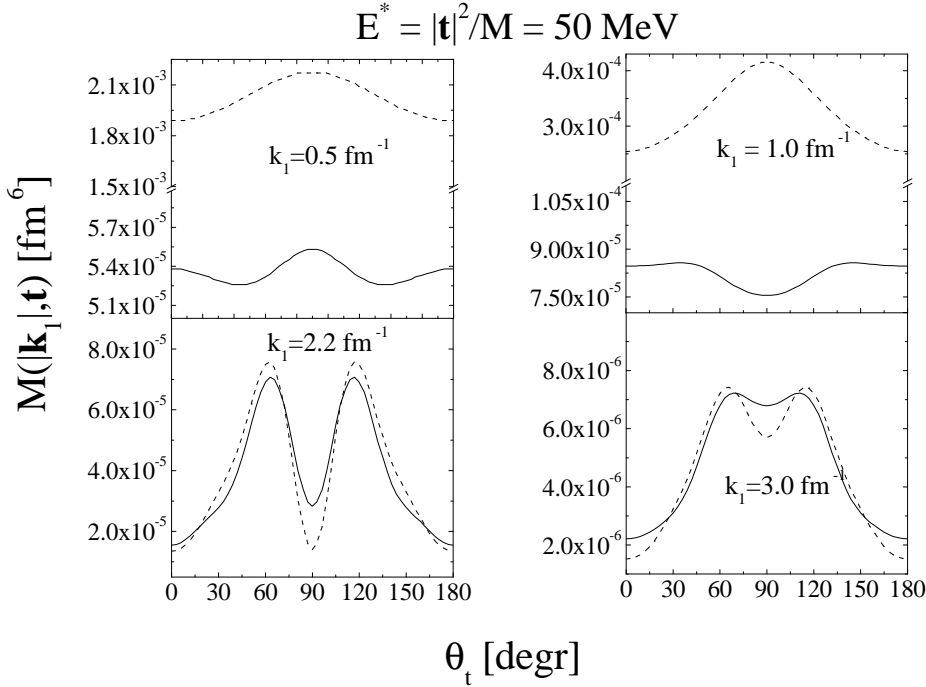}
\caption{The quantity $\enne$ (\ref{ten})
calculated at fixed value of $E^*=50\ MeV$,
versus the angle $\theta_t$ between the relative momentum
of the emitted nucleons ${\bf t}$ and the momentum transfer ${\bf q}$. The full line
 includes the two nucleon rescattering and the dashed line
 represents the {\it No FSI} result, i.e. the three-body wave function
 in momentum space. The three values of $k_1$ which have been chosen, correspond to
 four  different regions of the Spectral Function (see text). Three-nucleon
 wave function from \cite{rosati}; AV18 interaction \cite{AV18}.}
\label{fig5}
\end{center}
\end{figure}

\begin{figure}[t]
\vskip -5mm
\raggedright\begin{minipage}[h]{75mm}
    \raggedright{\hspace*{-1mm}\includegraphics[height=0.38\textheight]{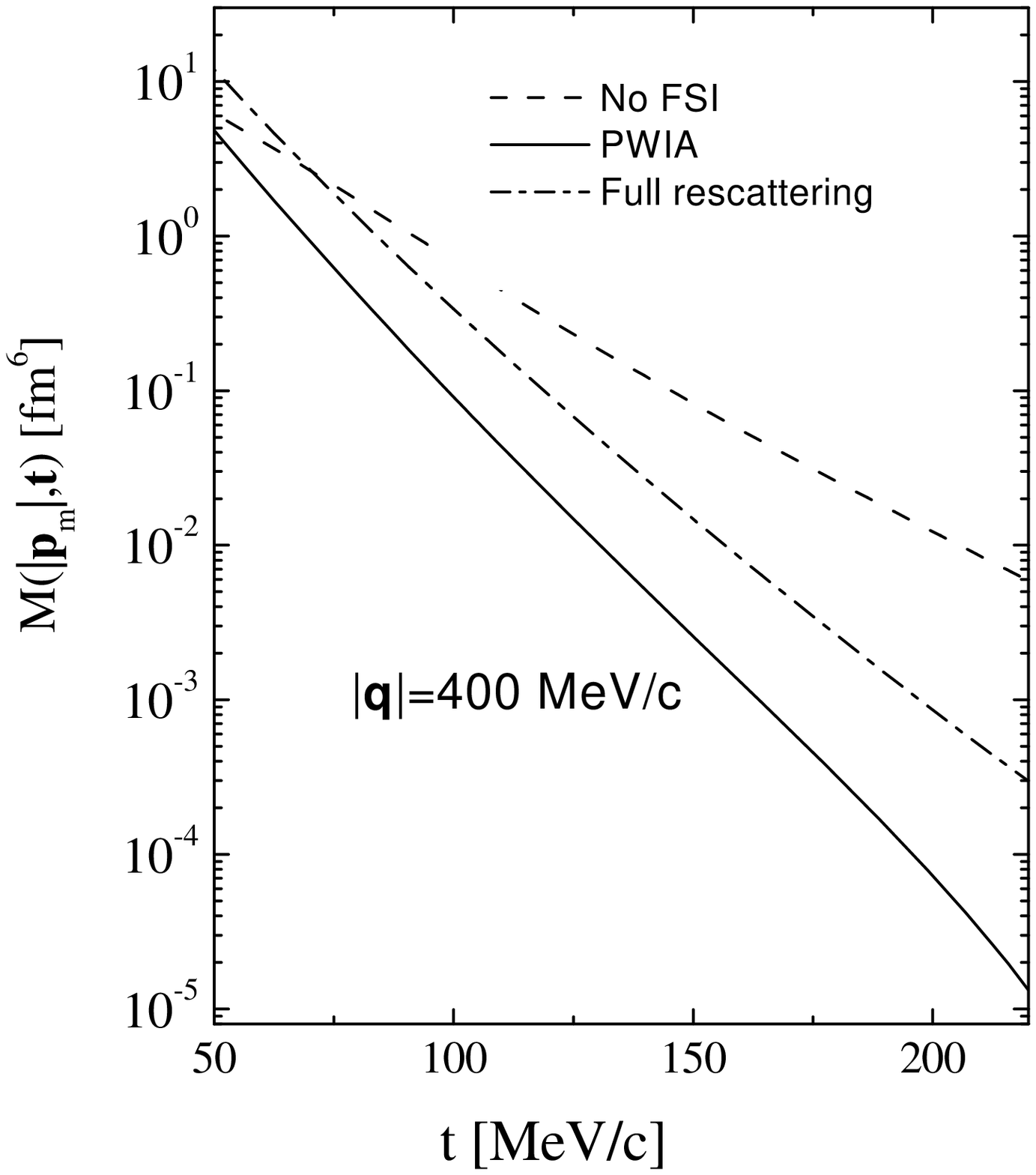}}
\caption{The quantity $M^D({\bf p}_m, {\bf t})$ (Eq. (\ref{twenty}),
 dot-dashed line)  calculated with ${\bf p}_2=-{\bf p}_3$ and
${\bf p}_m=0$, which for the processes a) and b)  corresponds to
the kinematics where, in the initial state, nucleons $N_2$ and $N_3$ were correlated
 with momenta ${\bf k}_2=-{\bf k}_3$ and ${\bf k}_1=0$. The dashed line
  corresponds
 to the {\it No FSI} case (Eq.\ref{4teen}), whereas the full line
 includes the {\it two nucleon rescattering}(Eq. \ref{ten}).
Three-nucleon
 wave function from \cite{rosati}; AV18 interaction \cite{AV18}.}
\label{fig6}
\end{minipage}
\hspace{\fill}
\begin{minipage}[h]{70mm}
   \vskip -7mm
    \hspace*{-.1cm}\includegraphics[height=0.39\textheight]{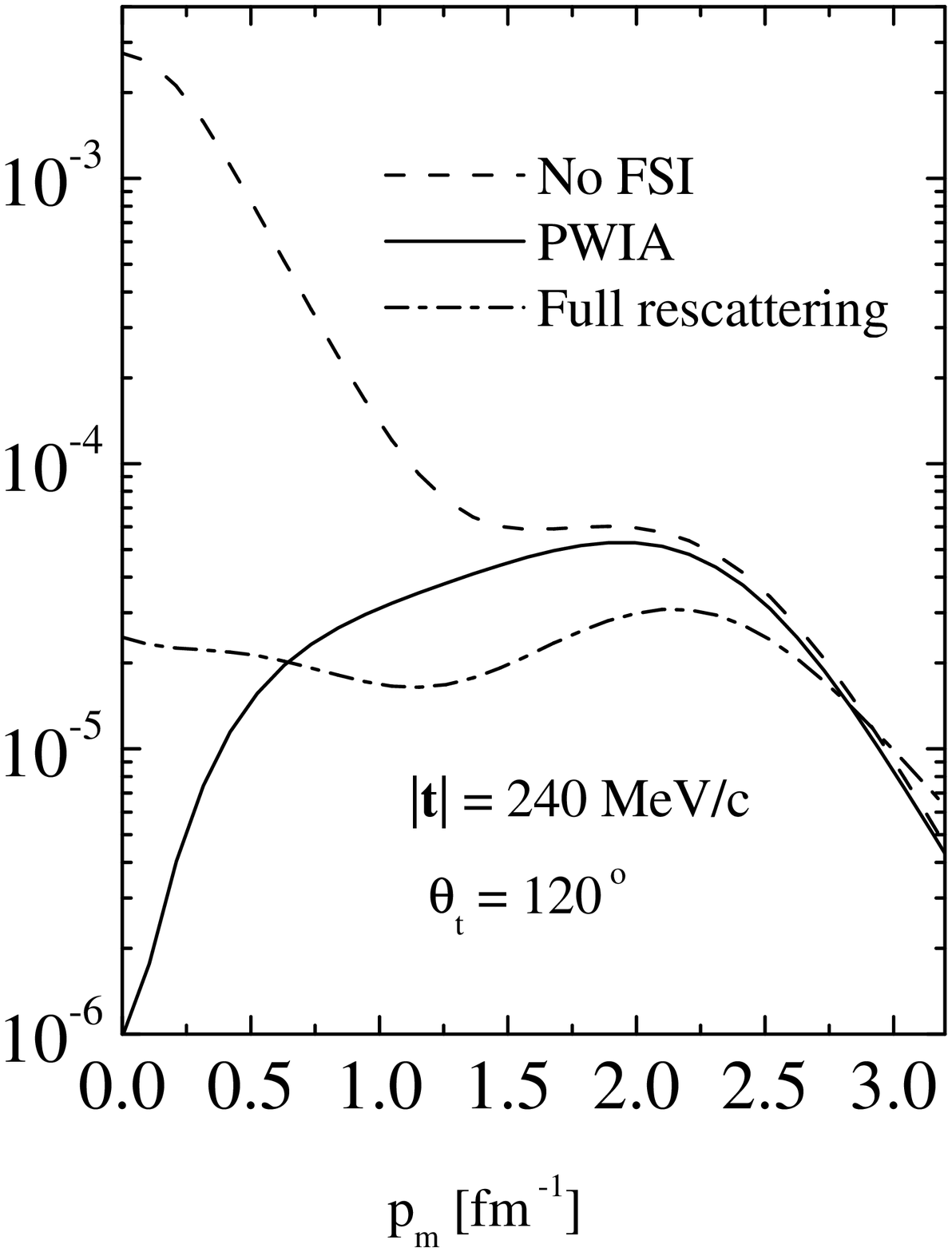}
    \caption{\label{fig7}
     The quantity $M^D({\bf p}_m, {\bf t})$ (Eq. (\ref{twenty}), dot-dashed line)
     calculated at fixed value of the relative momentum $|{\bf t}|=\, 240\,MeV/c$,
versus the missing momentum $p_m$. The full line
 includes the {\it two nucleon rescattering} (Eq.\ref{4teen}),
   whereas the dashed line
 represents the {\it No FSI} result (Eq. \ref{ten}). Three-nucleon
 wave function from \cite{rosati}; AV18 interaction \cite{AV18}.
      }
\end{minipage}
\end{figure}

The results are presented in Fig. \ref{fig6},  where  it can be seen that,
 as expected, the effect of
 the {\it NN rescattering}  is large. We have repeated this calculation in the correlated
region and found, obviously,
 that the rescattering, in this case,  has negligible effects \cite{leonya}.

We have eventually considered the three-body rescattering, e.g. process $c)$ of Fig. \ref{fig2}, by
treating the rescattering of $N_1$ with the interacting pair  $N_2N_3$ within an extended Glauber-type
approach \cite{glauber}. The details of the calculation will be presented elsewhere \cite{leonya}.

In Figs.\ref{fig6} and  \ref{fig7} we show our {\it preliminary results} for the quantity (dot-dashed line)
\be
M^D({\bf p}_m, {\bf t})=\left|\int \Phi_{N_1N_2N_3}^{{\bf p}_m} ({\bf r},{\bfgr \rho})
\, I_{N_2N_3}^{\bf t}( {\bfgr \rho}) d^3r d^3\rho \right|^2
 \label{twenty}
 \ee
  which is the generalization of Eq.(\ref{ten}) to take into account,
  {\it via} the quantity
 $\Phi_{N_1N_2N_3}^{{\bf p}_m} ({\bf r},{\bfgr \rho}) $,  the rescattering of $N_1$ with the
 interacting pair ${N_1N_2}$.
 In the Figure, $M^D({\bf p}_m, {\bf t})$ is plotted {\it vs} the
 missing momentum ( ${\bf p}_m \neq {\bf k}_1$ ) for a fixed value of ${\bf t}$;
 in the same Figure we also show   the results
 corresponding to the case when only the {\it NN rescattering} is active (full line) and to the case when
 all FSI's are  switched off (dashed line). In this calculation
  we have considered
 high values of $|{\bf q}|$, such that the asymptotic values of those quantities which enter
 the calculation (e.g. the total NN cross section, the ratio of the imaginary to  real
 parts of the forward scattering amplitude, etc.) have been adopted.
A more correct calculation, along the line of Ref. \cite{glauber},  will be presented
elsewhere \cite{leonya}.  It can be seen in Fig. \ref{fig7} that
in the {\it two-nucleon correlation} region and   at  high values of the missing
 momentum, the full FSI merely reduces to a change of the amplitude,
 without appreciably
   distorting
  the missing momentum distributions calculated without any
  type of FSI. Such a  result appears to be a very promising one in the
  investigation of  the correlated part of the three-body
 wave function.

 \section{Summary}
 We have investigated the effects of the Final State Interaction in the process
 of two-nucleon emission off $^3He$ induced by medium energy electrons.
  To this end, we have used
 realistic  three-nucleon wave functions, which, being the exact solution of the
 Schroedinger equation, incorporate all types of correlations, in particular the
  short-range and tensor  ones
 generated by modern NN potentials. Using these wave functions, we have
 investigated the process
 $^3He(e,e'pp)n$ ($^3He(e,e'np)p$) which can be generated
 by two main mechanisms: i)the absorption of $\gamma^*$ by  a non correlated nucleon,
  followed by the emission of  two "high"  momenta  nucleons, namely
  the ones  which were correlated
  in the initial state, and ii) the absorption of $\gamma^*$ by
  a nucleon of a correlated pair, followed  by the emission of a
  "high" momentum nucleon, which was the second correlated nucleon in the
  initial state, and a "low" momentum nucleon, corresponding to the spectator
  one in the initial state. A very specific kinematics corresponding
  to case $i)$, with the active nucleon  at rest in the initial
  state leaving the nucleus with momentum ${\bf p}_1$=${\bf q}$,
   followed by the back-to-back emission of the two correlated
  nucleons with momenta ${\bf p}_2$=-${\bf p}_3$,
   was also analyzed. We have taken into account the final
  state interaction both between the two detected nucleons, as well as
  between these and the active nucleon which absorbed $\gamma^*$.
    We have investigated the above process in different kinematical
  regions governed by various types of correlations. The so called
   {\it two-nucleon correlation} region, leading to process $i)$, turns
   out to be of particular interest, for, in such a region,
 Final State Interaction effects can be minimized

\end{document}